\documentclass[aps,a4paper, preprint, superscriptaddress, preprintnumbers, floatfix,nofootinbib,amsmath,amssymb]{revtex4}
\usepackage{url}
\usepackage{hyperref}
\usepackage{cleveref}

\usepackage{amstext,amssymb}
\usepackage{amsmath}
\usepackage{amsfonts}
\usepackage{amssymb}
\usepackage{graphicx}
\usepackage{xspace}
\usepackage{color}
\usepackage{units}

\usepackage{cancel}
\usepackage{bm}
\usepackage{slashed} 
\usepackage{multirow}
\newcommand{\be}{\begin{equation}}
\newcommand{\ee}{\end{equation}}
\newcommand{\bea}{\begin{eqnarray}}
\newcommand{\eea}{\end{eqnarray}}
\newcommand{\nn}{\nonumber}

\def\s1{\hat s}

\usepackage{slashed}

\newcommand{\nua}[1]{\ensuremath{\rlap{\kern-2.5pt\ensuremath{\overset{\scriptscriptstyle(-)}{\phantom{\nu}}}}{\ensuremath{{\nu}_{#1}}}}\xspace}

\begin{document}
\title{Vector leptoquark $U_3$ and CP violation at T2K, NOvA experiments}
\author{ Rudra Majhi}
\email{rudra.majhi95@gmail.com}
\affiliation{School of Physics,  University of Hyderabad, Hyderabad - 500046,  India}
\author{ Dinesh Kumar Singha}
\email{dinesh.sin.187@gmail.com}
\affiliation{School of Physics,  University of Hyderabad, Hyderabad - 500046,  India}
\author{K. N. Deepthi}
\email{nagadeepthi.kuchibhatla@mahindrauniversity.edu.in}
\affiliation{Department of Physics, \'Ecole Centrale School of Engineering - Mahindra University, Hyderabad, Telangana, 500043, India}
\author{ Rukmani Mohanta}
\email{rmsp@uohyd.ac.in}
\affiliation{School of Physics,  University of Hyderabad, Hyderabad - 500046,  India}

\begin{abstract}
In the current epoch of neutrino physics, many experiments are aiming for precision measurements of oscillation parameters. Thus, various new physics scenarios which alter the neutrino oscillation probabilities in matter deserve careful investigation. In this context, we study the effect of a vector leptoquark which induces non-standard neutrino interactions (NSI) that modify the oscillation probabilities of neutrinos in matter. 
We show that such interactions provide a relatively large value of NSI  parameter $\varepsilon_{e \mu}$. Considering this NSI parameter, we successfully explain the recent discrepancy between the observed $\delta_{CP}$ results of T2K and NOvA. 

\end{abstract}

\maketitle
\flushbottom

\section{INTRODUCTION}
\label{sec:intro}
Existence of neutrino oscillation \cite{Osc-1,Osc-2,Osc-3}  in nature itself suggests that Standard Model of particle physics is incomplete and  it gives a hint towards physics beyond the standard model.
The elusive nature of neutrinos along with their least understood mass generation  mechanism indicate that neutrino sector is indeed a very exciting and promising sector, which plays a crucial role in exploring the nature of new physics. Precise measurement of neutrino oscillation parameters is the prime objective of the currently running and future neutrino oscillation experiments, which leads to the advancement of neutrino physics.

Apart from standard oscillation theory, non-standard neutrino interactions (NSIs) appear as sub-leading effect that influence the neutrino oscillation probability \cite{NSI-1,NSI-2,NSI-3, NSI-4,NSI-5}. There are myriad of studies in the literature, exploring the effects  of  NSIs on neutrino oscillation parameters, see e.g.,  \cite{NSI-6} and references therein. Atmospheric neutrinos are very sensitive to matter NSIs as they travel a very long distance through earth matter core. The exploration of NSIs with atmospheric data has been investigated in \cite{Fornengo:2001pm,Gonzalez-Garcia:2004pka,Friedland:2004ah}. Super-Kamiokande collaboration has studied the NSIs with atmospheric neutrino data \cite{Super-Kamiokande:2011dam}. Search for matter induced NSIs  have been performed with  current and future long-baseline accelerator experiments MINOS \cite{Kopp:2010qt,Mann:2010jz,Friedland:2006pi,Blennow:2007pu,Isvan:2011fa}, K2K \cite{Friedland:2005vy}, OPERA \cite{Ota:2002na,Esteban-Pretel:2008ztv,Blennow:2008ym}, T2K \cite{Adhikari:2012vc} and NOvA \cite{Friedland:2012tq}. MINOS and OPERA collaborations have put bound on $\epsilon_{\mu \tau}$ parameter at 90$\%$ C.L.\cite{Neutrino:2012,Blennow:2008ym}. The MiniBooNE short baseline experiment  has also explored matter NSIs \cite{Johnson:2006rs,Akhmedov:2010vy}. A combined study of Superbeam and reactor experiment in the presence of NSIs has been done in \cite{Kopp:2007ne}, while for reactor data only, it has been done in \cite{Ohlsson:2008gx}.

NSI parameters may affect to the measurement of oscillation parameters and unknowns in the neutrino sector. 
It should be emphasized that, the measurement of leptonic CP violation is one of the foremost goals of neutrino oscillation experiments. The T2K data released in 2020 has indicated that $\delta_{CP} \sim 3\pi/2$  \cite{ADunne, T2K-1} along with a preference for normal mass ordering. On the other hand, the data from currently running NOvA experiment
\cite{AHimmel, NOVA-1} suggested that $\delta_{CP} \sim \pi$. There is a mild tension at $2\sigma$ level between the measurements of these two experiments. This could be because of the limited statistics or systematic errors or more interestingly it could be a first hand hint for a new physics scenario.  Recently, two independent groups have shown in their studies that presence of $\epsilon_{e \mu}$ and $\epsilon_{e \tau}$ can resolve the current NOvA and T2K tension in the measurement of CP violating phase $\delta_{C P}$ \cite{Peter-denton,SSChatterjee}. In \cite{Rahaman-1}, it has been shown that this tension arises primarily from the  $\nu_e$ appearance data, while a review on the possible role of new physics leading to this tension has been discussed in \cite{Rahaman-2}.
 
 In this paper, we consider a model-dependent approach to investigate the effect of NSIs on neutrino oscillation. In particular,  envisage the leptoquark (LQ) model, where the new interactions are generated due to the exchange of a vector leptoquark  $U_3$  between the neutrinos propagating through the earth and the nucleons in the earth matter. We show that these interactions can give rise to large non-standard interaction (NSI) parameter $\varepsilon_{e \mu}$.  Further, we estimate the LQ couplings through which we obtain the values of off-diagonal NSI parameter $\varepsilon_{e\mu}$. We show that in the presence of this complex parameter the discrepancy in the $\delta_{CP}$ value of T2K and NOvA will cease to exist. 
 
 The outline of the paper is as follows. In Sec \ref{Th-framework}, we provide the theoretical framework of vector leptoquark model and show how the leptoquark couplings are related to the NSI parameters of the neutrino sector. We then obtain the constraints on these parameters by considering the branching  fraction of $\pi^0 \to \mu^\pm e^\mp$.
 We also present a brief discussion on the constraints on leptoquark coupling parameters obtained from lepton flavour violating muon decay modes: $\mu \to e \gamma$  and $\mu \to ee e$.
 In Sec \ref{NOVA-T2K-tension}, we address the present discrepancy between T2K and NOvA on the measurement of the CP violating phase $\delta_{CP}$, due to the presence of NSI. The discussion on the prediction of the neutrino oscillation parameters is presented in Sec \ref{Predictions}. The implication of $U_3$ leptoquark is briefly discussed in  Sec \ref{LFV-decays}, followed by the Conclusion in Sec \ref{Conclusion}.
 

\section{Theoretical Framework}
\label{Th-framework}
\begin{figure}[htb!]
\includegraphics[scale=0.75]{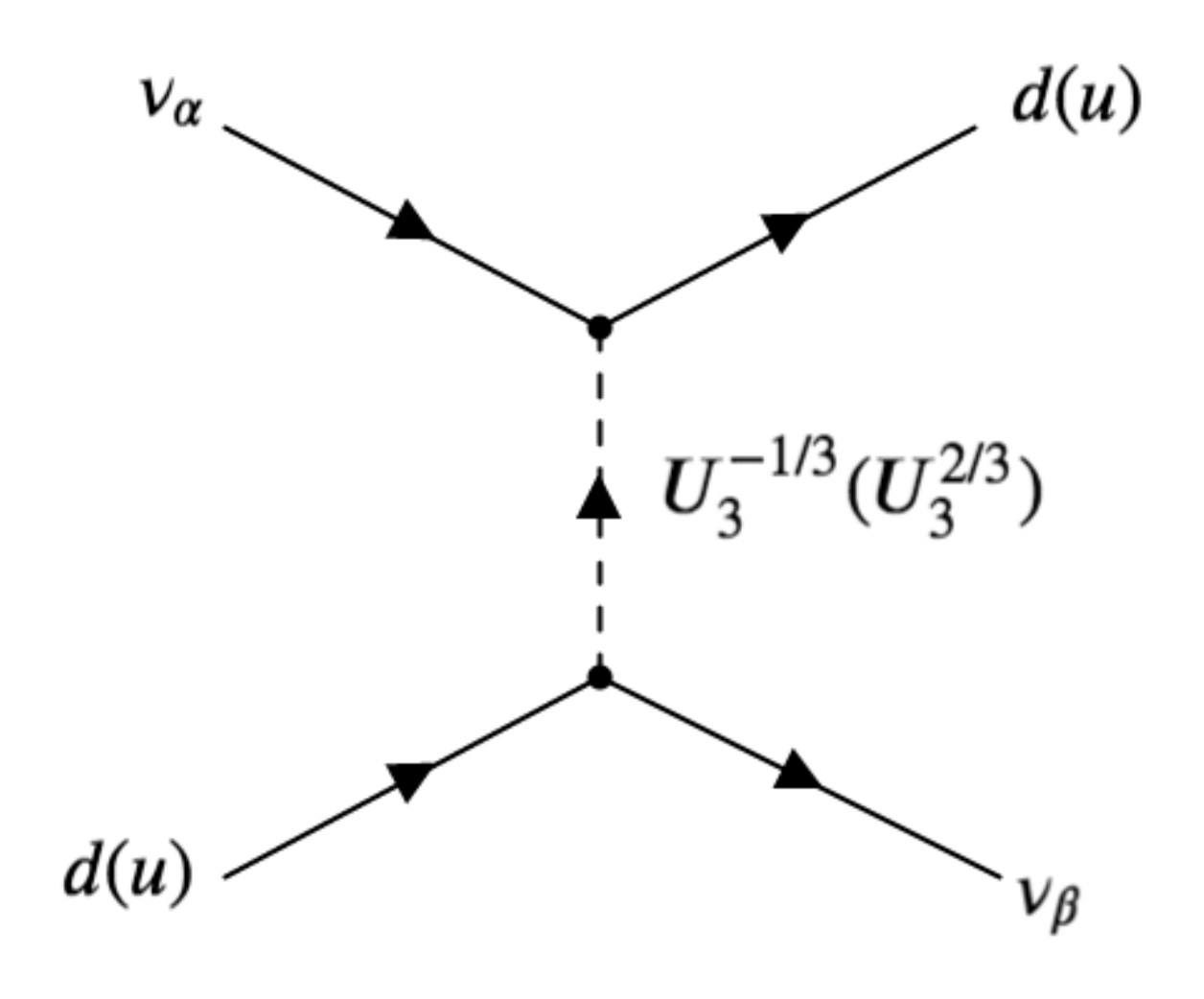}
\caption{Feynman diagram for the interaction of neutrinos with $u$ and $d$ quarks through the exchange of $U_3$ vector leptoquark.}
\label{fig:feyn-diag1}
\end{figure}

Without going into the details of any model building aspect, here we consider an additional vector leptoquark $U_3$ in addition to the SM particle content, which transforms as $(\overline{ 3},3,2/3)$ under the SM gauge group $SU(3)_C \times SU(2)_L \times U(1)_Y$. The vector LQ $U_3(\overline{ 3},3,2/3)$ can successfully accommodate the recently observed lepton flavour universality violating (LFV) observables: $R_{K^{(*)}}={\rm  BR}( B \to K^{(*)} \mu \mu)/{\rm BR}(B \to K^{(*)} ee)$, associated with the flavour changing neutral current transitions $b \to s \ell^+ \ell^-$, as illustrated  in Refs. \cite{LQ-12, Kosnik, RM}.   Furthermore, it does not couple to diquarks and hence,  conserves Baryon and Lepton numbers and does not induce proton decay.
   The existence of LQs at low energy is predicted in several extensions of the SM, e.g.,  Grand unified theory (GUT) \cite{LQ-1, LQ-2, LQ-3, LQ-4}, Pati-Salam model \cite{LQ-5,LQ-6,LQ-7}, technicolor \cite{LQ-8,LQ-9,LQ-10}, composite model \cite{LQ-11} etc. Since the LQs can couple to quarks and leptons simultaneously, they can induce new interactions between the propagating neutrinos with the nucleons in the earth matter.
The   charge of the  vector LQ $U_3(\overline{ 3},3,2/3)$ is expressed in terms of its  isospin and hypercharge as $Q=I_3+Y$ and thus, the three charged states are denoted as $U_3^{5/3}, U_3^{2/3}, U_3^{-1/3}$. During the propagation, the  neutrinos interact with the nucleons (i.e. $u$ and $d$ quarks)  of the earth matter through the exchange  of the leptoquark as shown in the Figure-\ref{fig:feyn-diag1}.

Since the vector leptoquark $U_3$ transforms as a triplet under $SU(2)_L$, it can couple only to left-handed quark and lepton doublets and the corresponding  Lagrangian describing the interaction  is given as \cite{fajfer}
\bea
{\cal L} \supset \lambda_{ij}^{LL}~
 \overline Q_L^{ i,a}  \gamma^\mu  ( \tau^k \cdot U_{3,\mu}^k)^{ab} L_L^{j,b}+{\rm H.c.}, \label{eq:Lag}
 \eea 
where $Q_L(L_L)$ are the left-handed quark (lepton) doublets,    $\boldsymbol{\tau}$ are the Pauli matrices, $i,j = 1,2,3 ~(a,b = 1,2)$ are the generation $ (SU(2))$ indices.  $U_3^k$ are components of $U_3$ in $SU(2)$ space and the charged states are related to them as $U_3^{5/3}=(U_3^1-iU_3^2)/\sqrt{2}$, $U_3^{-1/3}=(U_3^1+iU_3^2)/\sqrt{2}$ and $U_3^{2/3}=U_3^3$. 
The  convention used here to assign the chirality in the LQ couplings $\lambda^{LL}_{ij}$ is that the quark chirality precedes that of the lepton. It should be noted that the $U_3$ leptoquark does not  couple to diquark and hence,  it does not induce proton decay. 
Expanding the flavour indices, one can write the interaction Lagrangian (\ref{eq:Lag}) in the flavor basis as 
\bea
{\cal L} \supset -\lambda_{ij}^{LL}~
 \overline d_L^{ i}  \gamma^\mu  U_\mu^{2/3} e_L^j+\lambda_{ij}^{LL}  \bar u_L^i \gamma^\mu U_\mu^{2/3} \nu_L^j 
+\sqrt{2} \lambda_{ij}^{LL} \overline d_L^i \gamma^\mu U_\mu^{-1/3} \nu_L^j + \sqrt 2 \lambda_{ij}^{LL}  \overline u_L^i \gamma^\mu U_\mu^{5/3} e_L^j   +{\rm H.c.}, \hspace*{0.2 true cm}
\label{eq:Lag1}
 \eea 
 with $\psi_{L,R}= P_{L,R} \psi$, where $P_{L,R}=(1 \mp \gamma_5)/2$ are the chiral projection operators.
 
The transition from flavor to mass basis can be accomplished through the following transformations: $u_L^j \to V^\dagger_{jk}u_L^k$,  $d_L^j \to d_L^j$, $e_L^j \to e_L^j$ and  $\nu_L^j \to U^\dagger_{jk}\nu_L^k$, where $U$ and $V$ represent  the PMNS and CKM mixing matrices, and it is assumed that  the flavor basis coincides with the mass basis for the down-type quarks and charged leptons.

Thus, one can write the effective four-fermion interaction term between the  neutrinos and {\it up} or {\it down}-type quarks, i.e., for the interaction of the type  $(q^i + \nu_\alpha \to q^j+ \nu_\beta )$,  ($\alpha, \beta=e,\mu,\tau)$ as
\bea
&&{\cal L}_{\rm eff}^{\rm down}= -\frac{2}{m_{LQ}^2} \lambda^{LL}_{j \beta}{\lambda^{LL}_{i \alpha}}^* (\overline d^i \gamma_\mu P_L d^j)(\bar \nu_\alpha \gamma^\mu P_L \nu_\beta)\;,\nn\\
&&{\cal L}_{\rm eff}^{\rm up}= -\frac{1}{m_{LQ}^2} \lambda^{LL}_{j \beta}{\lambda^{LL}_{i \alpha}}^*(\overline u^i \gamma_\mu P_L u^j)(\bar \nu_\alpha \gamma^\mu P_L \nu_\beta)\;,
\label{eq:Lag2}
\eea
where $m_{LQ}$ represents the mass of the leptoquark.

Recalling the standard expression for the neutral current non-standard interaction Lagrangian \cite{NSI-5}, which is generally represented by
\bea
{\cal L}_{\rm NSI} = - 2 \sqrt 2 G_F \varepsilon_{\alpha \beta}^{f C} (\overline \nu_\alpha \gamma^\mu P_L \nu_\beta)(\overline f \gamma_\mu P_C f)\;,
\label{eq:Lag3}
\eea
 where $\alpha, \beta=e,\mu,\tau$ indicate the neutrino flavor, $f=e,u,d$ represent the matter fermions, the superscript $C=L,R$ refers to the chirality of the fermion
 current and $\varepsilon_{\alpha \beta}^{fC}$ are the strengths of NSIs. The Hermiticity of the interaction implies
 \bea
 \varepsilon_{\beta \alpha}^{fC} = (\varepsilon_{\alpha \beta}^{fC})^*.
 \eea
 The NSI contributions which are 
relevant as neutrino propagate through the earth are those coming from the interaction of neutrino with $e$, $u$ and $d$ because the 
earth matter is made up of these fermions only. Therefore, the effective NSI parameter is given by
\begin{equation}
\varepsilon_{\alpha\beta} = \sum_{f=e,u,d}\frac{n_f}{n_e} \varepsilon_{\alpha\beta}^{f} \,,
\end{equation}
where $\varepsilon^f_{\alpha\beta}= \varepsilon_{\alpha\beta}^{fL} + \varepsilon_{\alpha\beta} ^{fR}$, and $n_f$ ($n_e$)  represents the number density of the fermion $f$  (electrons)  in earth. For earth matter, we can assume that the number densities of electrons, protons and neutrons are equal, i.e., 
$n_n \approx n_p =n_e$, in which case $n_u \approx n_d = 3 n_e$. Therefore, one can write $\varepsilon_{\alpha \beta}$ as 
\bea
\varepsilon_{\alpha \beta} \simeq \varepsilon_{\alpha \beta}^{e} + 3\varepsilon_{\alpha \beta}^{u}+ 
3\varepsilon_{\alpha \beta}^{d}\;.
\label{eq:bd}
\eea
 
 Thus, comparing eqns. (\ref{eq:Lag2}) and (\ref{eq:Lag3}), we get the relation between the NSI parameter $\varepsilon_{\alpha \beta}$ and LQ parameters as
 \bea
 \varepsilon_{\alpha \beta}^{uL} =\frac{1}{2 \sqrt 2 G_F}\frac{1}{m_{LQ}^2}\lambda_{1\beta}^{LL} {\lambda_{1\alpha}^{LL}}^*,~~~~~~~{\rm and}~~~~~~
 \varepsilon_{\alpha \beta}^{dL} =\frac{1}{ \sqrt 2  G_F}\frac{1}{m_{LQ}^2}\lambda_{1\beta}^{LL} {\lambda_{1\alpha}^{LL}}^* .\label{NSI-bnd}
 \eea
 
In this work,  we are mainly interested to study the effect of the NSI parameter $\varepsilon_{e \mu}$, which may be responsible for  the observed discrepancy between the NOvA and T2K results on the CP violating phase $\delta_{\rm CP}$.
Hence, we need to know the values of the LQ couplings $(\lambda_{1\mu}^{LL} {\lambda_{1e}^{LL}}^*)/m_{LQ}^2 \equiv (\lambda_{12}^{LL} {\lambda_{11}^{LL}}^*)/m_{LQ}^2$. 
For constraining these parameters, we consider the lepton flavour violating decays $\pi^0 \to \mu e$ and by confronting its predicted  branching fraction in the LQ model with the present available data we obtain the bound on the LQ couplings.

\begin{figure}[htb!]
\includegraphics[scale=0.75]{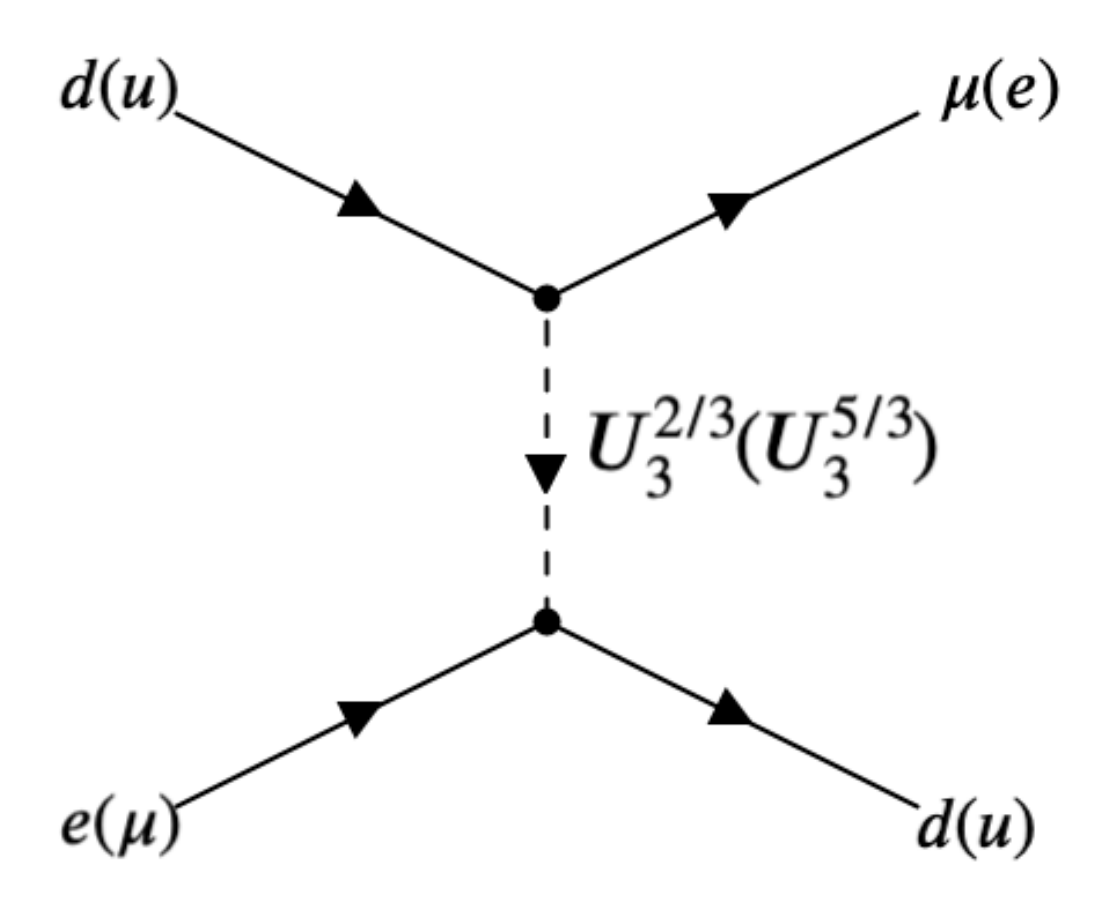}
\caption{Feynman diagram for the decay of $\pi^0 \to \mu^\pm e^\mp$ through the exchange of $U_3$ LQ.}
\label{fig:feyn-diag2}
\end{figure}

The lepton flavor violating process $\pi^0 \to \mu^\pm e^\mp $ is mediated through the exchange of $U_3^{2/3} (U_3^{5/3})$ LQ as shown in Fig. \ref{fig:feyn-diag2}.
The effective Lagrangian  for $\pi^0 \to (\mu^+ e^-+e^+ \mu^-)$ process is given as 
\bea
&&{\cal L}_{\rm eff}= -  \Big [ \frac{1}{ m_{LQ}^2}\lambda_{12}^{LL} {\lambda_{11}^{LL}}^*(\bar d_L \gamma^\mu d_L) (\bar \mu_L \gamma_\mu e_L)  \nn\\
&&~~~~~~+ \frac{2}{m_{LQ}^2}(V \lambda^{LL})_{12} {(V\lambda^{LL})}_{11}^*(\bar u_L \gamma^\mu u_L) (\bar \mu_L \gamma_\mu e_L)  \Big]+{\rm H.c.},
\eea
where the LQ mass $m_{LQ}$ is  assumed to be same for both the multiplets and $V$ is the CKM matrix. As the CKM matrix elements are strongly hierarchical, i.e., $V_{11}>V_{12}>V_{13}$, keeping only the diagonal element $V_{11}$, the branching fraction of $\pi^0 \to \mu e$ process is given as  
\bea
&&{\rm BR}(\pi^0 \to \mu^+e^- +\mu^- e^+) = \frac{1}{64 \pi m_\pi^3}
\frac{\left |\lambda_{12}^{LL}{ \lambda_{11}^{LL}}^* \right |^2}{m_{LQ}^4} \tau_\pi f_\pi^2
\left(1- 2 V_{11}^2\right )^2 \nn\\
&&~~~~~~~~~\times \sqrt{(m_\pi^2 -m_\mu^2 -m_e^2)^2 -4 m_\mu^2 m_e^2}~
 \Big[ m_\pi^2(m_\mu^2+m_e^2) - (m_\mu^2-m_e^2)^2 \Big].
\eea
 The  measured branching ratio of this process is given in terms of 90\% C.L. upper limit (UL) as ${\rm BR}(\pi^0 \to \mu^+ e^-+\mu^-e^+) < 3.6 \times 10^{-10}$ \cite{pdg}.  For constraining the LQ parameters, we convert the UL to a branching ratio of $0\pm {\rm UL}/1.5$. 
Using the particle masses and lifetime of $\pi$ meson  $f_\pi=0.130$ GeV from \cite{pdg}, we obtain the bound on the leptoquark parameters as
\bea
0\leq \frac{|\lambda_{12}^{LL} {\lambda_{11}^{LL}}^*|}{m_{LQ}^2} \leq  3.4 \times 10^{-6}~{\rm GeV^{-2}}.
\eea
Considering the LQ couplings to be  ${\cal O}(1)$, this gives a lower bound on the LQ mass as $m_{LQ} \geq 540$ GeV. These bounds can be translated into NSI couplings using  Eqn (\ref{NSI-bnd}) as
\bea
\varepsilon_{e \mu}^{uL} \leq 0.1,~~~~~~~\varepsilon_{e \mu}^{dL} \leq  0.2,
\eea
which gives $\varepsilon_{e \mu} \leq 0.9$.

Now, we briefly discuss the effect of the Non-Standard Interactions of the neutrinos with the earth matter during their   propagation.  Due to the presence of NSIs, the effective Hamiltonian of neutrinos gets modified and is represented as 
\be 
{\cal H} = \frac1{2E}\left[U M^2 U^\dagger + V_m\right]\,,
\label{eq:nsi_hamil}
\ee
where $M^2=\text{diag}(0,\Delta m_{21}^2,\Delta m_{31}^2)$ and $U$ is the PMNS mixing matrix parametrized by three mixing angles $ \theta_{12},\theta_{13},\theta_{23} $ and a CP phase $ \delta_{CP} $. 
Here, $V_m$ represents the matter potential due to the standard charged-current interactions (the unit term in the ${(V_m)}_{11}$ element of the matrix) and non-standard neutral current interactions of neutrinos,
\be
V_m = A \left(\begin{array}{ccc}
1 + \varepsilon_{ee} & \varepsilon_{e\mu}e^{i\phi_{e\mu}} & \varepsilon_{e\tau}e^{i\phi_{e\tau}}
\\
\varepsilon_{e\mu}e^{-i\phi_{e\mu}} & \varepsilon_{\mu\mu} & \varepsilon_{\mu\tau}e^{i\phi_{\mu\tau}}
\\
\varepsilon_{e\tau}e^{-i\phi_{e\tau}}& \varepsilon_{\mu\tau}e^{-i\phi_{\mu\tau}} & \varepsilon_{\tau\tau}
\end{array}\right)\,,
\label{eq:potential}
\ee
where,
$A \equiv 2\sqrt2 G_F n_e E$, $n_e$ stands for the density of electrons. 

Using matter perturbation theory, the appearance probability $P_{\mu e}$ to first order in $A$, can be expressed in terms of the off-diagonal $\varepsilon_{e\mu}$ for normal mass hierarchy (NH) as \cite{Minakata}
\begin{eqnarray}
P_{\mu e} &=& 
P_{\mu e}(\varepsilon=0)_{SI} +
P_{\mu e}(\varepsilon_{e \mu})_{NSI}, 
\label{Pmue-mattP}
\end{eqnarray}
where the standard interactions term (SI) is called Arafune-Koike-Sato formula and is expressed as
\begin{eqnarray}
P_{\mu e}(\varepsilon=0)_{SI} &=& 
\sin^2{2\theta_{13}} s^2_{23} \sin^2 \Delta_{31} + 
c^2_{23} \sin^2{2\theta_{12}} 
\left( \frac{ \Delta m^2_{21} }{ \Delta m^2_{31} } \right)^2 
\Delta_{31}^2 
\nonumber \\
&+& 
4 c_{12} s_{12} c_{13}^2 s_{13} c_{23} s_{23} \left( \frac{ \Delta m^2_{21} }{ \Delta m^2_{31} } \right)
\Delta_{31}
\left[
\cos{\delta}
\sin 2 \Delta_{31} - 
2 \sin{\delta}
\sin^2 \Delta_{31} 
\right] 
\nonumber \\
&+& 
2 \sin^2{2\theta_{13}} s^2_{23} 
\left( \frac{AL}{4 E} \right)
\left[
\frac{1}{ \Delta_{31} }
\sin^2 \Delta_{31} 
- \frac{1}{2} \sin 2 \Delta_{31} 
\right],
\label{AKS}
\end{eqnarray}
while the contribution arising from the NSIs is given as
\begin{eqnarray}
&&P_{\mu e}(\varepsilon_{e \mu} \neq 0)_{NSI} = - 8 \left( \frac{AL}{4E} \right) 
\nonumber \\
&\times& 
\left[
s_{23} s_{13} 
\left\{ 
\vert \varepsilon_{e \mu} \vert \cos (\delta + \phi_{e \mu} )
\left( 
s^2_{23} \frac{ \sin^2 \Delta_{31} }{ \Delta_{31} } - 
\frac{c^2_{23} }{2} \sin 2\Delta_{31}
\right) 
+ c^2_{23} \vert \varepsilon_{e \mu} \vert \sin (\delta + \phi_{e \mu} )
\sin^2 \Delta_{31} 
\right\}
\right.
\nonumber \\
&&\hspace*{-6mm} {} -
\left.
c_{12} s_{12} c_{23} 
\frac{ \Delta m^2_{21} }{ \Delta m^2_{31} } 
\left\{
\vert \varepsilon_{e \mu} \vert \cos \phi_{e \mu} 
\left( c^2_{23} \Delta_{31} + \frac{s^2_{23} }{2} \sin 2\Delta_{31} \right) 
+ 
s^2_{23} 
\vert \varepsilon_{e \mu} \vert \sin \phi_{e \mu} \sin^2 \Delta_{31} 
\right\} 
\right],
\label{P-NSI-emu}
\end{eqnarray}
where $\Delta_{31} \equiv \frac{ \Delta m^2_{31} L} {4 E} $.

\section{Addressing the T2K, NOvA discrepancy of CP violating phase $\delta_{CP}$}
\label{NOVA-T2K-tension}
The general approach to test the standard three flavor neutrino oscillation paradigm is to accurately measure the neutrino oscillation probabilities and compare them with the theoretical predictions. If there is any mismatch in the observed data and the predictions,  it could mean that there exists new physics beyond the  standard three flavor neutrino picture. In this regard, it is worthwhile to explore the reason behind the tension between the recent T2K and NOvA measurements of $\delta_{CP}$. The primary and obvious difference between these two experiments is the difference in their baseline lengths. NOvA has a 810 km baseline while T2K has a comparatively shorter baseline of 295 km. Therefore, one possibility is to attribute this tension to the neutrino-nucleus interactions in earth matter during the propagation of the beam. 

Recently,  it has been shown in Ref~{\cite{Peter-denton}} that this mismatch in the $\delta_{CP}$ data of T2K and NOvA can be sorted in the presence of complex non-standard interaction (NSI) parameters affecting the neutrino propagation. In Ref~{\cite{SSChatterjee}}, the authors have shown that the existence of the flavor changing NSI parameters $\varepsilon_{e\mu}$ and $\varepsilon_{e\tau}$ can successfully explain the discrepancy in the observed $\delta_{CP}$ value by NOvA and T2K.
In this work, we study the effect of non-zero NSI parameter $\varepsilon_{e\mu}$ as obtained from the vector leptoquark ($U_3$) induced interactions between neutrinos and nucleons. Thus, we consider all NSI parameters except $\varepsilon_{e\mu}$ to be zero.

When we consider the presence of non-zero $\varepsilon_{e \mu}$, one can find degenerate solutions for Eq.~(\ref{Pmue-mattP}) with two different set of oscillation parameters, i.e., 
\begin{eqnarray}
 P_{\mu e}(\theta_{12}, \theta_{13}, \theta_{23},  \delta_{CP}^{\rm true},  \Delta m^2_{21}, \Delta m^2_{31}, \varepsilon_{e \mu} ,\phi_{e \mu})_{\rm NSI}=P_{\mu e}(\theta_{12}, \theta_{13}, \theta_{23},  \delta_{CP}^{\rm meas},  \Delta m^2_{21}, \Delta m^2_{31})_{\rm SI} .\hspace*{0.4 true cm}
\label{Pmue-deg}
\end{eqnarray}

From eqns. (\ref{AKS}) and (\ref{P-NSI-emu}), one can obtain a relationship between the measured and true values of $\delta_{CP}$ for the NOvA experiment, after performing a little algebraic manipulation with normal ordering as \cite{Peter-denton},
\begin{eqnarray}
&&-s_{12}c_{12}c_{23} \frac{\pi}{2}\sin \delta_{CP}^{\rm true}+A |\varepsilon_{e \mu}| \Big(s_{23}^2\cos( \delta_{CP}^{\rm true} +\phi_{e \mu})-c_{23}^2 \frac{\pi}{2} \sin(\delta_{CP}^{\rm true}+\phi_{e \mu} )\Big)\nonumber\\
&&\approx -s_{12}c_{12}c_{23} \frac{\pi}{2}\sin \delta_{CP}^{\rm meas}\equiv P_{\mu e}^\delta\;. \label{P-del}
\end{eqnarray}
\begin{figure}[htb!]
\includegraphics[height=6cm,width=9cm]{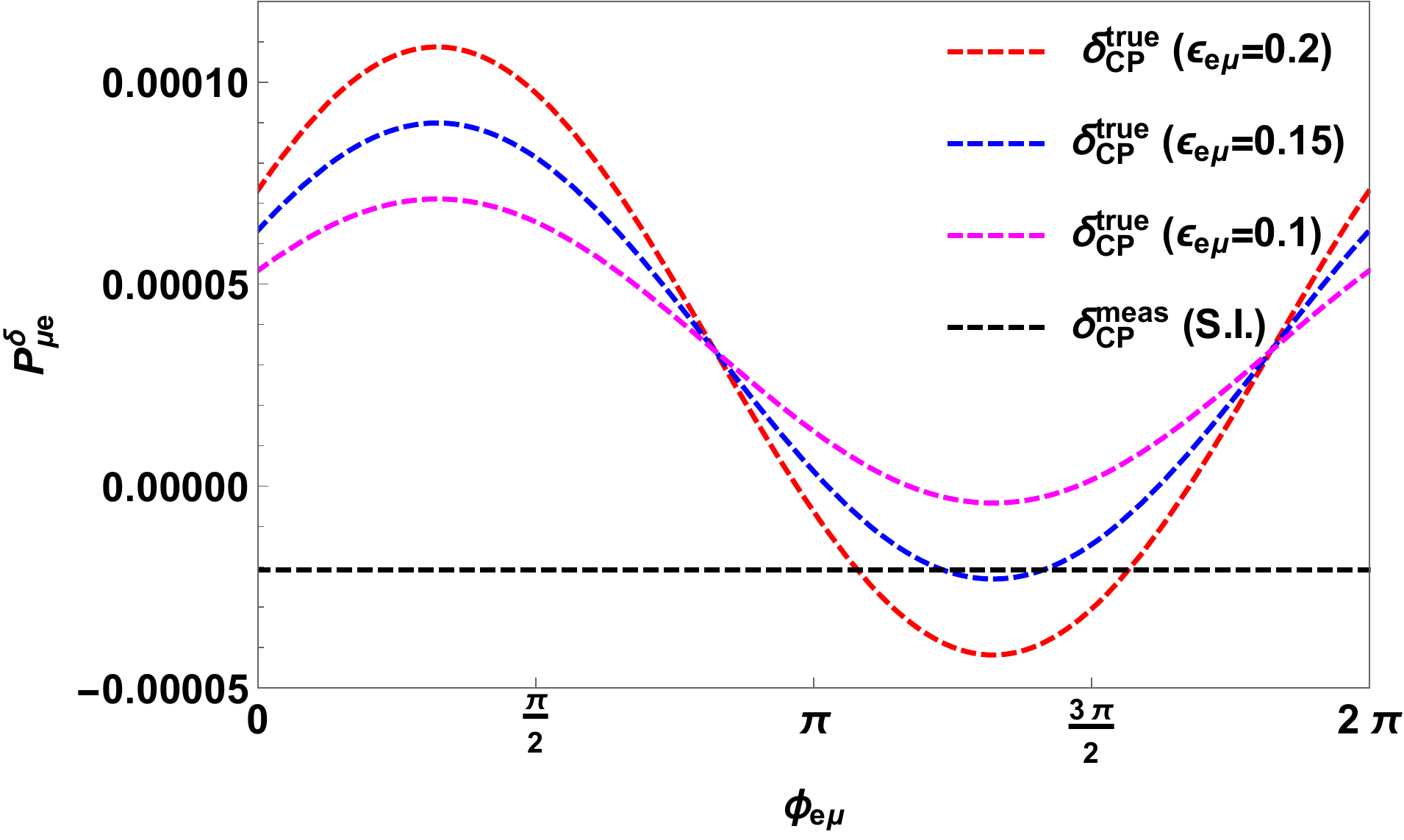} 
\caption{Appearance probability $P_{\mu e}$ versus $\phi_{e \mu}$ assuming standard interactions (black curve) and NSI for different values of $\varepsilon_{e \mu}$.}
\label{fig:prob-nsi}
\end{figure}  
Fig.~\ref{fig:prob-nsi} shows the behaviour of $P_{\mu e}^\delta$ with respect to the phase $\phi_{e \mu}$ for NOvA experiment. This can be obtained by plotting the l.h.s and r.h.s of eqn (\ref{P-del}) separately.  The black curve shows the $\delta_{CP}$ part of the appearance probability $(P_{\mu e}^\delta)$ for standard neutrino interactions (SI). The red, blue and magenta curves are plotted assuming the presence of non-zero $\varepsilon_{e \mu}$ for   $\varepsilon_{e \mu}= 0.2$, $\varepsilon_{e \mu}= 0.15$ and $\varepsilon_{e \mu}= 0.1$ respectively. It can be seen that there is no intersection between magenta curve ($\varepsilon_{e \mu}= 0.1$) and the black curve (SI) indicating that there is no degeneracy between SI and NSI parameters for this value of $\varepsilon_{e \mu}$. On the other hand, when $\varepsilon_{e \mu}= 0.15$ (blue), $\varepsilon_{e \mu}= 0.2$ (red), we can clearly see an overlap between red, blue and black curves demonstrating a degeneracy of the form eq. (\ref{Pmue-deg}) for different values of $\phi_{e \mu}$. However, in the case of T2K experiment the effect of NSI at a baseline of 295 km is not very prominent. 
We show that the degeneracies of this form led to tension between the $\delta_{CP}$ measurement in NOvA and T2K. Therefore, we consider $\varepsilon_{e \mu}= 0.15$ as a representative value through out our analysis. This value of  $\varepsilon_{e\mu}$ corresponds to the leptoquark coupling parameters  as
\bea
\frac{|\lambda_{12}^{LL} {\lambda_{11}^{LL}}^*|}{m_{LQ}^2} =  5.5 \times 10^{-7}~{\rm GeV^{-2}},\label{eq:20}
\eea
which yields  $m_{LQ} \approx 1.3$  TeV for  ${\cal O}(1)$ leptoquark couplings. It should be emphasized that the recently observed  LFU violating $B$  anomalies ($R_{K^{(*)}}$) can be successfully explained by a TeV scale $U_3$ vector leptoquark \cite{LQ-12, Kosnik, RM}. 

\subsection*{Simulation Details}
We have used GLoBES software package \cite{globes1} along with some additional files \cite{globes-nsi1} for simulating T2K and NOvA experiments in our analysis. The experimental specifications are as given below:

$NOvA$ (NuMI off-axis $\nu_e$ Appearance): It is a long-baseline experiment with a near detector (ND) of fiducial volume 280 ton placed at a distance of 1 km and far detector (FD) of mass 14 kton placed at an off-axis angle of 14.6 mrad at 810 km away from the source \cite{NOVA}.  The ND observes the unoscillated $\nu_\mu$ beam and the FD observes $\nu_e$ oscillated beam. Primary objective of the experiment is to observe the $\nu_\mu$ to $\nu_e$ oscillation for both neutrino and anti neutrino cases. The NuMI beam of Fermilab provides high intensity $\nu_\mu$ beam of power 0.7 MW and protons of 120 GeV energy corresponding to $1.36 \times 10^{20}$ protons on target (POT) per year. The energy flux peaks at 2 GeV for the neutrino beam.

$T2K$ (Tokai to Kamioka) is a long baseline experiment with baseline length of 295 km and off axis angle $2.5^\circ$ \cite{T2K}. The $\nu_\mu$ beam from Tokai is first observed by ND at 280 meters from the source and the oscillated $\nu_e$ beam is detected by FD of fiducial mass 22.5 kton at Kamioka. The energy of the neutrino flux peaks at 0.6 GeV. The energy of the proton beam is 30 GeV with beam power of 750 kW, which corresponds to $7.8\times 10^{21}$ POT with neutrino and anti neutrino mode ratio as 1:1. 

In this work we simulated NOvA and T2K experiments according to Refs.~\cite{AHimmel} and \cite{ADunne} using GLoBES package. Later, we obtained the statistical $\chi^2$ using Poissonian $\chi^2_{{\rm}}$ function 
\begin{eqnarray}
 \chi^2_{{\rm stat}} = 2 \sum_i \Big\lbrace N_i^{{\rm te}}-N_i^{{\rm tr}}+N_i^{{\rm tr}} \ln \frac{N_i^{{\rm tr}}}{N_i^{{\rm te}}} \Big\rbrace, 
 \label{mh-eq}
\end{eqnarray}
where $N_i^{{\rm te}}$ corresponds to the number of events predicted by the model while $N_i^{{\rm tr}}$ denotes the total number of simulated events (signal and background) in $i^{th}$ bin. The systematic uncertainties are incorporated into the simulation using Pull method. {The true values of oscillation parameters used for our calculation are shown in Table~\ref{table_nsi1}.

\begin{table}
\centering
\begin{tabular}{ |p{3cm}|p{3cm}|p{3cm}|  }
\hline
Parameters& NOvA & T2K \\
\hline
$\sin^2 2\theta_{1 2}$& 0.851 & 0.851 \\
$\sin^2 2\theta_{1 3}$& 0.085 & 0.086  \\
$\sin^2 \theta_{2 3}$ & 0.57 & 0.546 \\
$\delta_{C P}$ & 1.4$\pi$ &  $1.4 \pi$\\
$\Delta m^{\rm 2}_{2 1}$ & 7.53$\times 10^{-5}$ ${\rm eV}^{2}$ & 7.53$\times 10^{-5}$ ${\rm eV}^{2}$ \\
$\Delta m^{\rm 2}_{3 1}$ & 2.41$\times 10^{-3}$ ${\rm eV}^{2}$ & 2.49$\times 10^{-3}$ ${\rm eV}^{2}$ \\
$\epsilon_{e \mu}$ & 0.15 & 0.15 \\
$\phi_{e \mu}$& 1.53$\pi$ & 1.53$\pi$ \\
\hline
\end{tabular}
\caption{True values of the oscillation parameters used in the analysis for NOvA and T2K. }
\label{table_nsi1}
\end{table}

\noindent \subsection*{Discussion at the Probability Level}

In Fig.~{\ref{fig:Osc-phi-emu}}, we show the oscillograms for appearance probability $P_{\mu e}$ assuming $\varepsilon_{e \mu} = 0.15$ with respect to the variation in $\delta_{CP}$ and the NSI phase $\phi_{e \mu}$. The left and the right plots correspond to NOvA and T2K experiments respectively. Here the red and blue colours represent the regions showing maximum and minimum probabilities respectively. In the case of T2K experiment, the maximum probability point is obtained for $\delta_{CP}=3\pi/2$ as indicated by the recent T2K data. Even in the case of NOvA experiment one can see from the left plot that the appearance probability is maximum around $\delta_{CP}=3\pi/2$. So, when one assumes the presence of non-zero $\varepsilon_{e \mu}$, both the experiments NOvA and T2K are suggesting same value of $\delta_{CP}=3\pi/2$.  
\begin{figure}[htb!]
\includegraphics[width=1\textwidth]{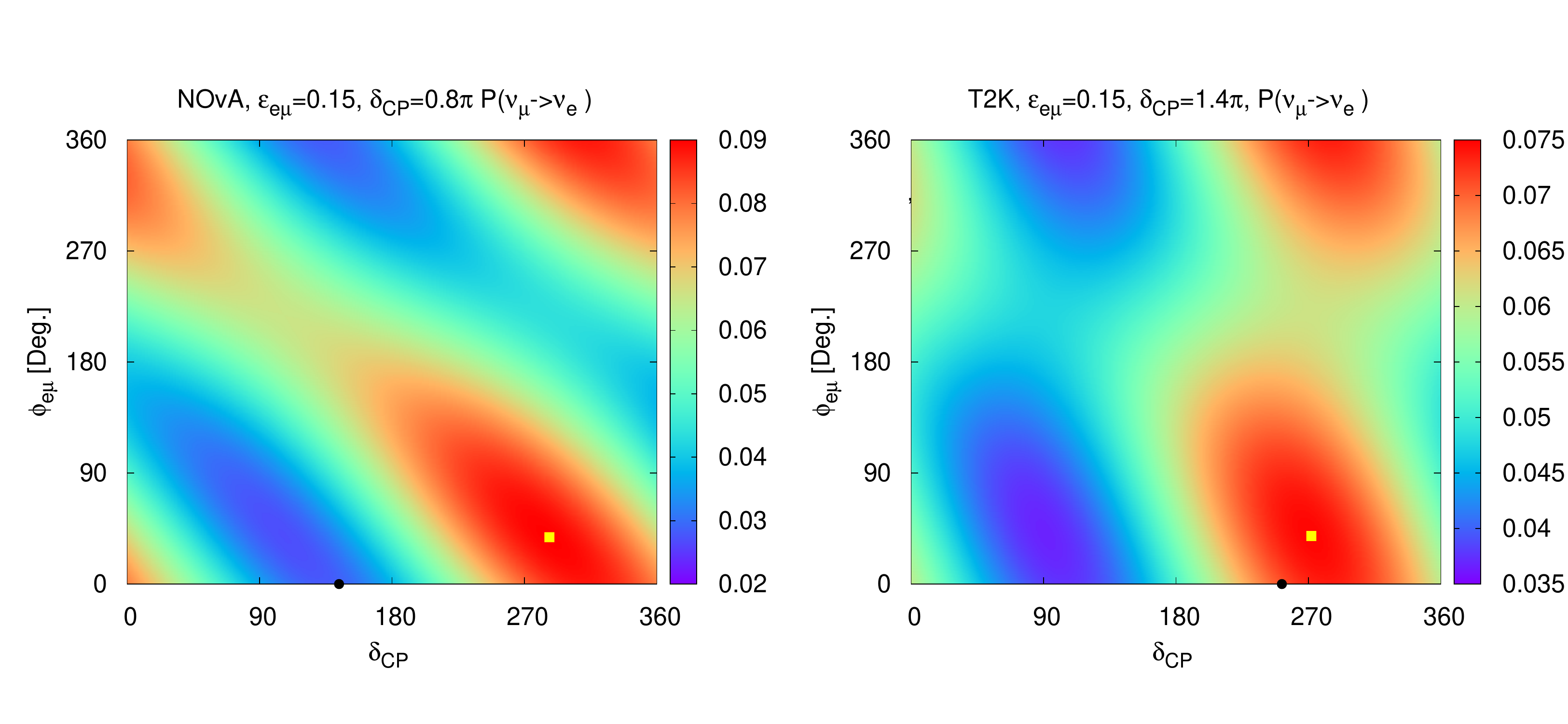}           
\caption{Oscillograms of $P_{\mu e}$ for $\varepsilon_{e \mu} = 0.15$, NH, plotted with respect to variation of standard CP phase $\delta_{CP}$ and the NSI phase $\phi_{e \mu}$. The left (right) plot corresponds to NOvA (T2K) baseline.}
\label{fig:Osc-phi-emu}
\end{figure}  
\section{Predictions for neutrino oscillation parameters}
\label{Predictions}
In this section, we explore the predictions of the model in the context of T2K, NOvA experiments.
Fig.~{\ref{fig:cont_emu_dcp} shows the allowed parameter regions spanned in the 
$\sin^2\theta_{23}$ - $\delta_{CP}$ plane. The red and green curves show the regions allowed at 1$\sigma$ and 2$\sigma$ confidence levels for 2 degrees of freedom. In both the cases we have assumed normal hierarchy and the presence of non-zero $\varepsilon_{e \mu}$. Additionally, we have marginalised over $\Delta m^2_{31}$ and $\phi_{e \mu}$. 
 The corresponding true values are listed in Table~\ref{table_nsi1}. 
  It can be seen from left and right plots that the allowed values of $\sin^2\theta_{23}$ and $\delta_{CP}$ spanned by the contours in both the experiments are in agreement with each other. We can observe from left plot that in the case of NOvA experiment the true point for $\delta_{CP}$ has been shifted from $0.8\pi$ to $1.5\pi$ thus, resolving the tension between the two experiments. 
\begin{figure}[htb!]
\includegraphics[height=6cm,width=8cm]{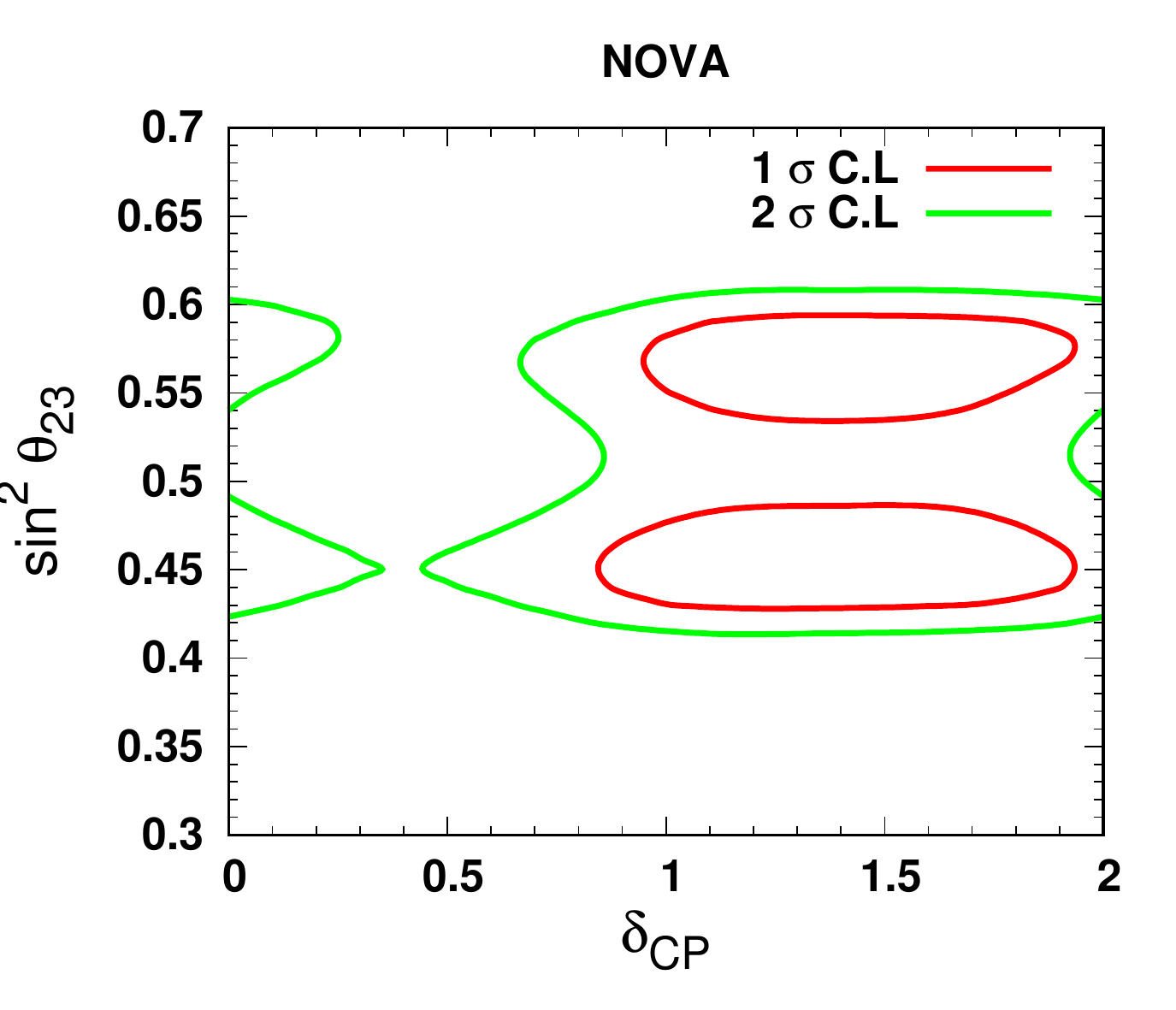} 
\includegraphics[height=6cm,width=8cm]{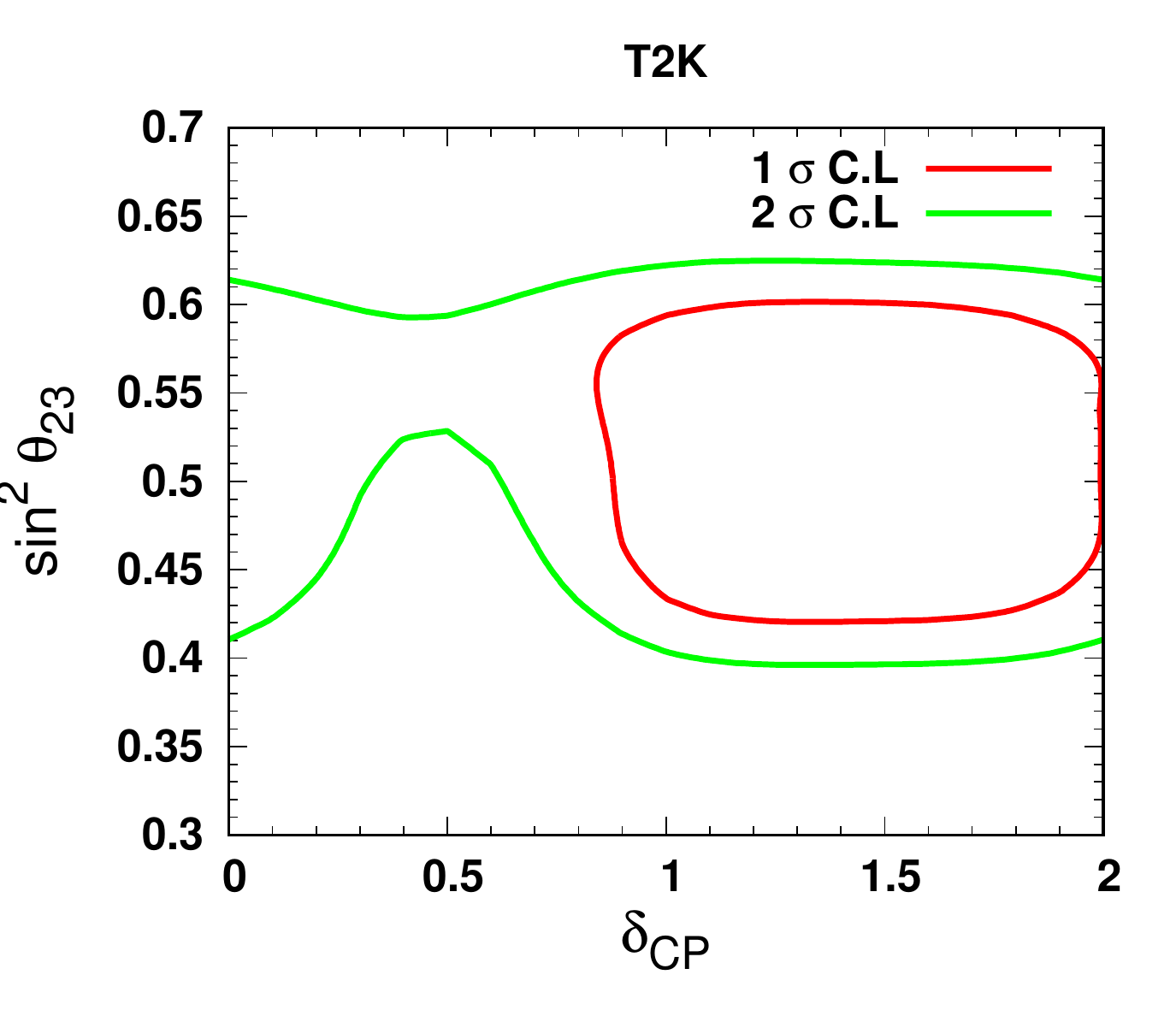} 
\caption{Allowed 1$\sigma$ and 2$\sigma$ confidence regions for $\sin^2\theta_{23}$-$\delta_{CP}$ parameter space obtained for NOvA (left) and T2K (right) experiments.}
\label{fig:cont_emu_dcp}
\end{figure}  
To further investigate the best fit value of $\theta_{23}$ in the presence of NSI for NOvA, we plot $\chi^2$ vs $\sin^2\theta_{23}$  in Fig~\ref{fig:chi2-sinth23}. Here the red and blue curves represent SI and NSI ($\varepsilon_{e \mu}=0.15$) respectively. The best fit value of $\theta_{23}$ in the case of standard interactions falls in higher octant as shown by the minima of the red curve. From the blue curve we can see that the absolute minima still falls in higher octant, though for values of $\theta_{23}<45^\circ$ there almost seems to be a degeneracy with lower octant.

\begin{figure}[htb!]
\includegraphics[scale=0.6]{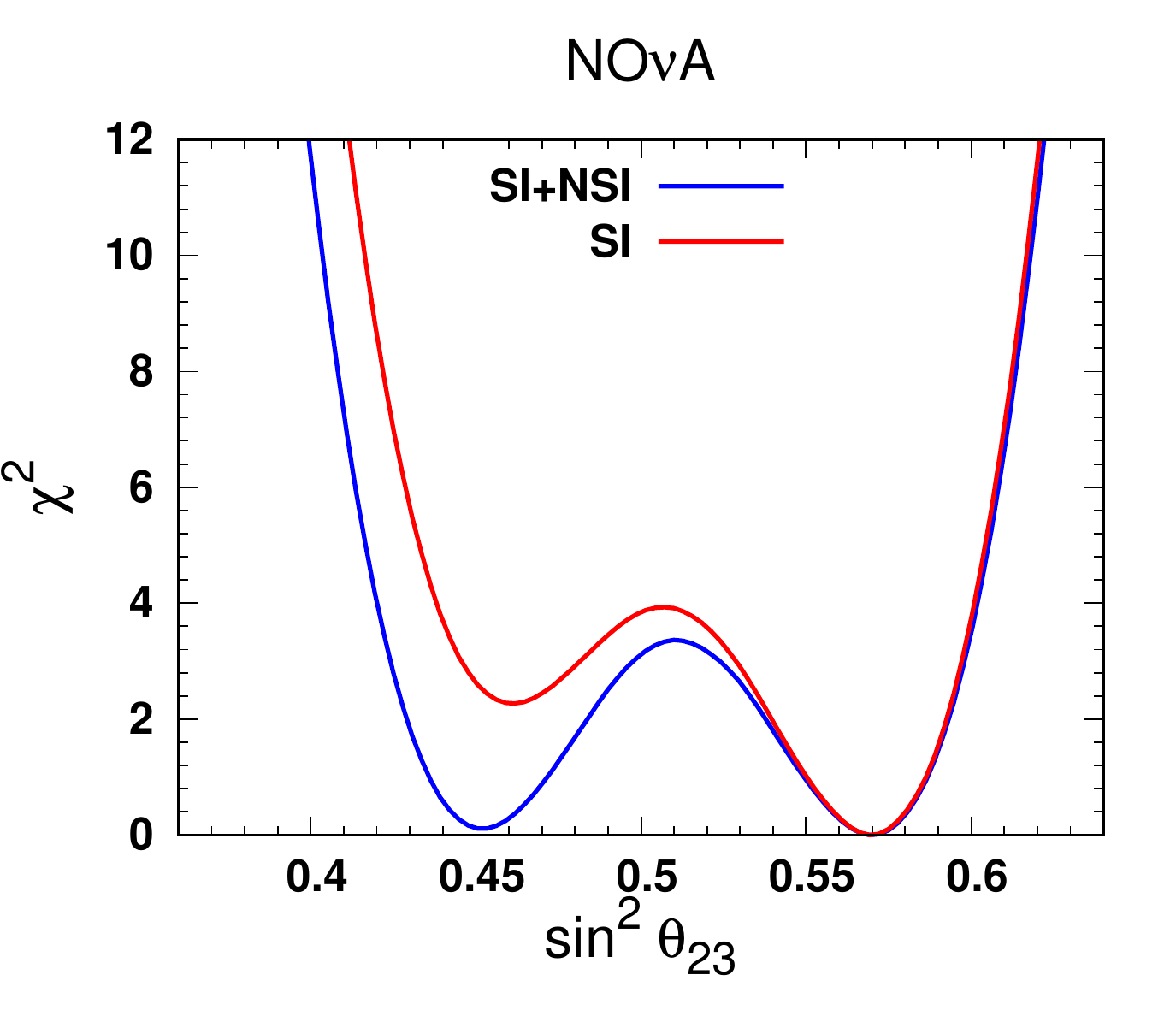}
\vspace{-0.5cm}
\caption{$\chi^2$ vs $\sin^2{\theta_{23}}$ for NOvA obtained assuming both SI (red) and NSI (blue).}
\label{fig:chi2-sinth23}
\end{figure}  
\section{ Implications of $U_3$ LQ on lepton flavour violating $\mu$ decays}
\label{LFV-decays}
 In this section, we will briefly discuss the implications of the $U_3$ leptoquark on the lepton flavour violating (LFV) rare $\mu$ decays, e.g., $ \mu \to e \gamma$ and $\mu^- \to e^- e^+ e^-$. The leptoquark couplings relevant for constraining the NSI parameter $\varepsilon_{e \mu}$ are intimately connected to these rare decay processes. 
 
The LFV decay mode $ \mu \to e \gamma$ is  strictly forbidden in the Standard Model and can be induced in models with extended gauge or lepton sectors. The current limit on the branching ratio is ${\rm BR}(\mu \to e \gamma) < 4.2 \times 10^{-13}$ at $90\%$ C.L. from MEG Collaboration \cite{MEG}. The general effective Lagrangian relevant for the  on-shell transition $\mu \to e \gamma $ is given as \cite{Lindner}
\bea
{\cal L}_{\rm eff} = \frac{\mu_{e \mu}^M}{2} ~(\bar e \sigma^{\mu \nu} \mu) F_{\mu \nu}+\frac{\mu_{e \mu}^E}{2} ~(\bar e i \gamma_5 \sigma^{\mu \nu} \mu)F_{\mu \nu}\;,\label{eq:22}
\eea
where $F_{\mu \nu}$ is the electromagnetic field strength tensor. Neglecting the mass of the electron and expressing  $\mu_{e \mu}^{M/E}=e m_\mu A_{e \mu}^{M/E}/2$, the branch ratio of $\mu \to e \gamma$ process is given as
\bea
{\rm BR}(\mu \to e \gamma)= \frac{3 (4 \pi)^3 \alpha }{4 G_F^2}\left (|A_{e \mu}^M|^2 +|A_{e \mu}^E|^2\right),\label{mu2eg}
\eea
where $\alpha$ represents the electromagnetic fine-structure constant.

In the presence the $U_3$ leptoquark, the $\mu \to e \gamma$ process can be mediated through one-loop  diagrams with $up/down$ quarks and $U^{2/3}/U^{5/3}$ leptoquarks flowing in the loop, as shown in Fig. \ref{fig:Feyn-5}. 
\begin{figure}[htb!]
\includegraphics[scale=0.75]{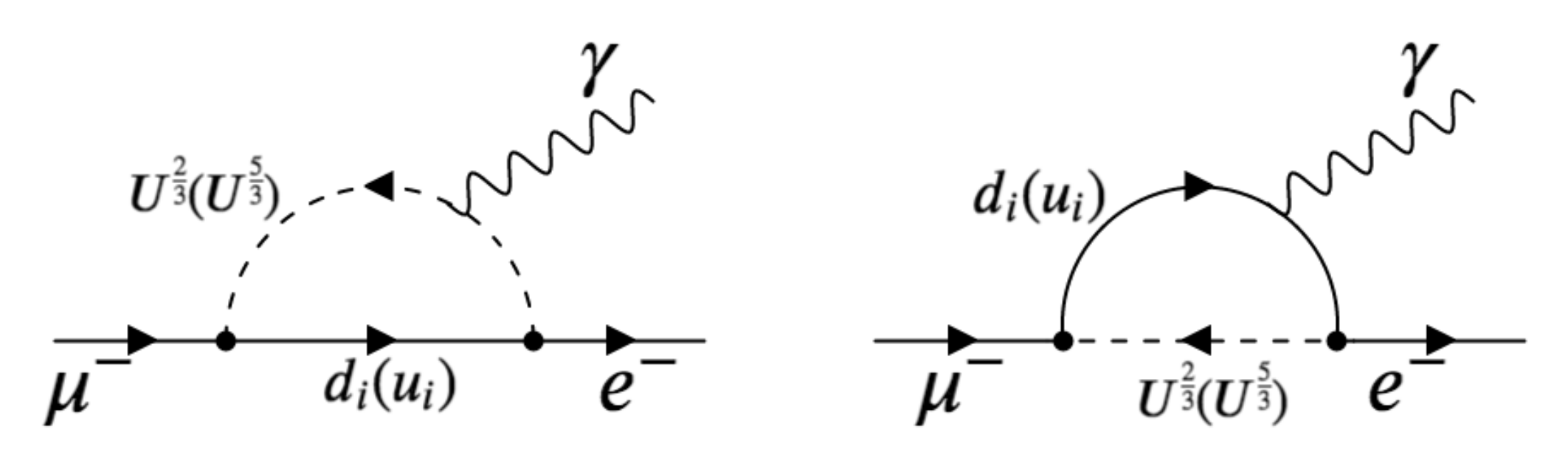}
\caption{Feynman diagrams for the LFV decay process $ \mu \to e \gamma$  through one-loop diagrams with $up/down$ type quarks and $U_3$ leptoquark flowing in the loop.}
\label{fig:Feyn-5}
\end{figure}

Thus, with eqns. \ref{eq:Lag}, \ref{eq:22} and \ref{mu2eg}, one can obtain the branching ratio for this process  in the presence of $U_3$ leptoquark,   including only the leading-order loop functions of order ${\cal O}(m_{q_i}^2/m_{LQ}^2)$ as \cite{Gab},
\bea
{\rm BR}(\mu \to e \gamma)=\frac{3 \alpha N_C^2}{64 \pi G_F^2} \Biggr [\sum_{i=1}^3 \frac{ |\lambda^{LL}_{i2} \lambda^{LL}_{i1}|}{m_{LQ}^2} \Big (\frac{1}{2}\frac{m_{d_i}^2}{m_{LQ}^2}  + \frac{m_{u_i}^2}{m_{LQ}^2} \Big )\Biggr ]^2,
\eea
where  $N_C=3$ is the number of colors and $m_{d_i}(m_{u_i})$ represent the masses of the $down (up)$-type quarks. The constraints  leptoquark couplings can be scrutinized by assuming the presence of  only one generation of quarks flowing in loop   at a given time. Thus, it should be noted that  $\mu \to e \gamma$ process is not very sensitive to the LQ couplings   $\lambda^{LL}_{12} \lambda^{LL}_{11}$, required for obtaining the limit on $\varepsilon_{e \mu}$, as $m_{(u/d)}^2/m_{LQ}^2$ is negligibly small.  
Now, using the value of leptoquark parameters from eqn. (\ref{eq:20}),  the quark masses as $m_u =m_d \approx 350 $ MeV, we obtain the branching ratio for a TeV scale leptoquark as, 
\bea
{\rm BR}(\mu \to e \gamma) \approx 7.35 \times 10^{-20}\;,
\eea
which is well below the present upper limit  ${\rm BR}(\mu \to e \gamma) < 4.2 \times 10^{-13}$, as expected.

\begin{figure}[htb!]
\includegraphics[scale=0.75]{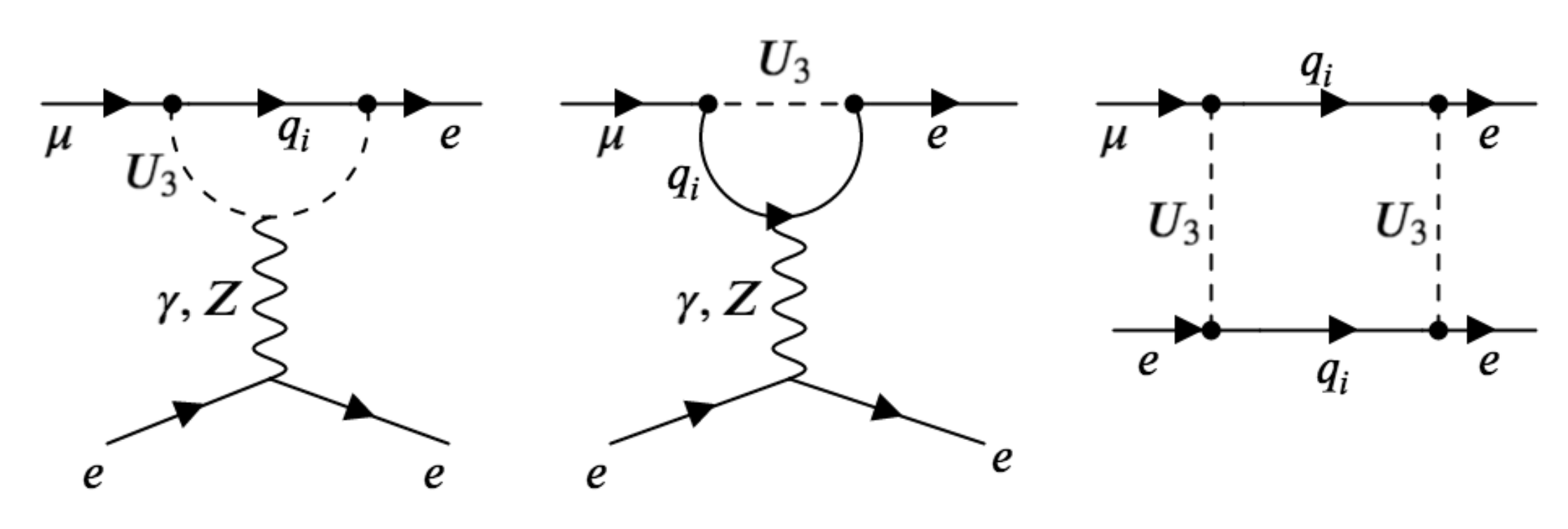}
\caption{Feynman diagrams for the LFV decay process $ \mu \to e ee $  through one-loop penguin and box diagrams  with $up/down$ type quarks and $U_3$ leptoquark flowing in the loop.}
\label{mu23e}
\end{figure}
Next, we consider the LFV decay mode $ \mu \to eee$, for which only the upper limit of  branching fraction exists, i.e., ${\rm BR}(\mu \to ee e) < 1.0 \times 10^{-12}$ \cite{pdg}.   The relevant penguin and box diagrams for this process are shown in Fig. \ref{mu23e}. Though this process is expected to be suppressed with respect to $\mu \to e \gamma$ process by a factor of $\alpha$, but in the limit of large $m_{LQ}$, its contribution is enhanced by ${\rm log}(m_q^2/m_{LQ}^2)$. 

The branching ratio for this  process in the presence of $U_3$ leptoquark is given  as \cite{Gab},
\bea
{\rm BR}(\mu \to eee)=\frac{\alpha^2 N_C^2}{96 \pi^2 G_F^2}\Biggr [\sum_{i=1}^3\frac{|\lambda^{LL}_{i2}\lambda^{LL}_{i1}|}{m_{LQ}^2} {\rm log} \left( \frac{m_{q_i}^2}{m_{LQ}^2}\right )\Biggr]^2.
\eea
As the branching fraction is proportional to $({\log}(m_{q_i}^2/m_{LQ}^2))^2$, it gets very large enhancement for the first generation of quarks in the loop and hence, suggests the corresponding  LQ couplings to be vanishingly small. Thus, for a TeV scale leptoquark, 
the value of the coupling strengths, constrained by the branching ratio ${\rm BR}(\mu \to ee e) < 1.0 \times 10^{-12}$,  are found to be  
\bea
\frac{|\lambda_{12}^{LL} {\lambda_{11}^{LL}}|}{m_{LQ}^2} < 1.0 \times 10^{-9}~{\rm GeV^{-2}}, \label{mu3e}
\eea 
which provide vanishingly  small value for $\varepsilon_{e \mu}$.


\section{Summary}
\label{Conclusion}
The recent measurements of the CP violating phase $\delta_{CP}$ by the two currently running long-baseline experiments NOvA and T2K are not in agreement with each other. NOvA prefers towards CP-conserving case with $\delta_{CP}^{\rm NOvA} \approx 0.8 \pi$, while T2K points towards maximal CP violation with the measured value $\delta_{CP}^{\rm T2K}\approx 1.4 \pi$. This tension, which is at the level of $2 \sigma$ hints towards the presence of some kind of new physics in neutrino oscillation beyond the three flavour paradigm. The simplest and obvious reason for accounting this discrepancy is the
 presence of non-standard neutrino interactions of neutrinos with the earth matter during their propagation. This speculation of NSIs looks reasonable as these two experiments differ in their baselines significantly and hence, the effect of NSI on NOvA could be more prominent than T2K.  This fact has been investigated in detail in Refs \cite{Peter-denton,SSChatterjee} and it has been shown that the NSI effects in the $e\mu$ and  $e \tau$ sector can successfully resolve this ambiguity for $\varepsilon_{e \mu} \approx 0.2 $ and  $\varepsilon_{e \tau} \approx 0.3 $.

There are various new physics scenarios which can provide such non-standard interactions for the neutrinos. In this work, we have considered the vector leptoquark model as an example and have shown that it can successfully resolve the observed discrepancy in the measurement of $\delta_{CP}$ by T2K and NOvA. The leptoquark couplings involved in the interaction between the neutrinos and the $up/down$ type quarks in the earth matter are constrained by considering the lepton flavour violating decay $\pi^0 \to e^\pm \mu^\mp$. 
As the branching fraction of this decay mode is not yet measured, we obtained the upper limit of the NSI parameter $\varepsilon_{e \mu} \leq 0.9$ by confronting the theoretical prediction with its current upper limit.  We then performed our analysis by considering $\varepsilon_{e \mu} =0.15$. We found that such a value of $\varepsilon_{e \mu}$ can successfully explain the observed discrepancy. We have further shown that the allowed regions in $\sin^2 \theta_{23}-\delta_{CP}$ plane in the presence of NSI for both T2K and NOvA. In addition, we also noticed that in the three flavour paradigm NOvA prefers upper octant for the atmospheric mixing angle, while in the presence of NSI there is a degeneracy between the  upper and lower octants. We also briefly discussed the implications of $U_3$  leptoquark on the lepton flavour violating muon decays, such as $\mu \to e \gamma$ and $\mu \to eee$. To summarize, if this discrepancy persists with larger statistics with improved data, it would provide an indirect evidence for the existence of vector leptoquark $U_3$. 

{\bf Acknowledgements} One of the authors (Rudra Majhi) would like to thank the DST-INSPIRE program for financial support.  DKS acknowledges  Prime Minister's Research Fellows program,  Govt. of India, for financial support. The work of RM is supported by Univ. of Hyderabad IoE project grant no. RC1-20-012.  We gratefully acknowledge the use of CMSD HPC  facility of Univ. of Hyderabad to carry out computations in this work.


\end{document}